\documentclass[a4paper,11pt]{article}
\pdfoutput=1

\usepackage{epstopdf} 
\usepackage[T1]{fontenc} 
\usepackage{lmodern} 
\usepackage[dvipsnames]{xcolor}

\usepackage{graphics}
\usepackage[utf8]{inputenc}
\usepackage{xspace}
\usepackage{amsmath}
\usepackage{amssymb}
\usepackage{url}
\usepackage{rotating}
\usepackage{enumerate}
\usepackage{graphicx}
\usepackage{epstopdf} 
\usepackage{pbox} 
\usepackage{pdflscape, afterpage, capt-of} 
\usepackage{slashed} 
\usepackage[normalem]{ulem}
\usepackage{nicefrac}
\usepackage[export]{adjustbox} 
\usepackage{makecell} 
\usepackage{subcaption} 
\usepackage{bbold}
\usepackage{stackengine} 
\usepackage{hyperref}
\usepackage{cite}

\usepackage{soul}

\usepackage{setspace}
\usepackage{def}

\newcommand{\cur}{\mathcal{J}}
\newcommand{\bvec}{\mathcal{A}}

\renewcommand{\W}{\mathcal{W}}
\newcommand{\eqrefeq}[1]{eq.$\,$(\ref{#1})}

\makeatletter
\def\hlinewd#1{
\noalign{\ifnum0=`}\fi\hrule \@height #1 \futurelet
\reserved@a\@xhline}
\makeatother

\begin{document} 

\title{\textbf{All-gluon amplitudes with off-shell recursion in
    multiplet bases}}

\date{}
\author{
  Oskar Bolinder\footnote{E-mail: \texttt{oskar@bolinder.com}}\,,
  Rikkert Frederix\footnote{E-mail: \texttt{rikkert.frederix@fysik.lu.se}}\;
  and Malin Sjodahl\footnote{E-mail: \texttt{malin.sjodahl@fysik.lu.se}}\vspace{5pt}
  \\
{\small\it Department of Physics, Lund University, Box 118, 221 00 Lund, Sweden}\\
}

\maketitle

\begin{abstract}
  \noindent The efficient computation of color-summed QCD amplitudes
  at high parton multiplicities remains a central challenge for
  precision collider predictions. Existing approaches using trace,
  color-flow, or adjoint bases suffer from non-orthogonality, which
  complicates the color algebra and scales poorly with multiplicity.
  In this work, we present an off-shell recursive framework for
  computing all-gluon tree-level amplitudes directly in orthogonal
  multiplet bases. Utilizing Wigner $6j$ coefficients, we construct
  an algorithm that
  builds multiplet-projected off-shell currents from lower-point currents.
  By optimizing the recursion through partial summation and caching,
  we find that the computational complexity of calculating $n$-gluon
  color-summed squared amplitudes scales as $\mathcal{O}(17^n)$. This
  demonstrates the potential competitiveness of multiplet bases for
  high-multiplicity processes.
\end{abstract}

\thispagestyle{empty}
\vfill

\newpage

\tableofcontents

\section{Introduction}
\label{sec:introduction}

The unprecedented luminosity and data output of the approved
High-Luminosity Large Hadron Collider (HL-LHC) upgrade pose enormous
challenges for precision predictions in QCD. Accurate modeling of
multijet final states, especially in the high-multiplicity regime,
demands efficient computations of partonic scattering
amplitudes. While the kinematic structure of tree-level QCD
amplitudes can be handled efficiently in the form of color-ordered or
dual amplitudes using recursive techniques, the associated color
algebra remains a significant computational challenge, particularly in
high-multiplicity regimes where the number of color configurations
grows rapidly.

Several types of bases exist for decomposing color structures in QCD,
including the well-known trace bases~\cite{Paton:1969je,Mangano:1987xk,Sjodahl:2009wx}, the Del
Duca–Dixon–Maltoni (DDM) bases~\cite{DelDuca:1999rs} (for all-gluon
amplitudes), and the color-flow bases~\cite{Maltoni:2002mq}. These
constructions are widely used in both analytic and numerical
implementations. However, they all suffer from a crucial drawback:
non-orthogonality of the basis vectors. As a result, the color algebra
often introduces a prohibitively large computation time, particularly
in the numerical evaluation of the squared matrix elements at high parton
multiplicities. One strategy to reduce the
cost of summing over all color configurations is color sampling,
where a finite number of color states are stochastically selected and
averaged over in the Monte Carlo
integration~\cite{Papadopoulos:2000tt,Mangano:2002ea,Gleisberg:2008fv,DeAngelis:2020rvq,Isaacson:2018zdi}. While
this reduces computational complexity, in practice it leads to an
increased variance in the evaluation of the phase-space integrals,
thereby requiring significantly more Monte Carlo samples to achieve a
given statistical precision.

To mitigate this, orthogonal bases constructed from irreducible
representations of the color gauge group---referred to as color
multiplet bases---have been introduced~\cite{Keppeler:2012ih,Sjodahl:2018cca,Alcock-Zeilinger:2016cva}. These
provide a framework in which the color-summed squared amplitudes
become straightforward to compute due to orthogonality.
Using multiplet bases, it is possible to perform the color decomposition
by using Wigner $3j$ and $6j$ coefficients (also known as symbols, or for short $3j$s and $6j$s), rather than by explicitly
constructing the actual color tensors, see e.g.~refs~\cite{Cvitanovic:2008zz,Sjodahl:2015qoa}.
Recent work~\cite{Alcock-Zeilinger:2022hrk,Keppeler:2023msu} has demonstrated
that a sufficient set of Wigner $6j$ coefficients 
can be analytically solved for, or recursively calculated. 
Due to their orthogonality, the use of multiplet bases is
particularly attractive for moderate to high-multiplicity applications
in QCD.

On the kinematic side, recursive
techniques~\cite{Berends:1987me,Caravaglios:1995cd,Caravaglios:1998yr,Draggiotis:1998gr,Britto:2004ap,Britto:2005fq,Cachazo:2004kj,Duhr:2006iq} 
have been instrumental in simplifying the computation of
amplitudes. These recursion relations build high-multiplicity
color-ordered or dual amplitudes\footnote{A notable exception is the
color-dressed recursion introduced in ref.~\cite{Duhr:2006iq}.} from
lower-multiplicity ones. However, when working with multiplet bases,
dual amplitudes are not the natural objects to compute; rather, it is 
natural to consider a method that directly yields the full amplitudes projected
onto orthogonal representations. In the case of
maximally helicity violating amplitudes~\cite{Parke:1986gb},
such a multiplet basis recursion~\cite{Du:2015apa} has been worked out based on
BCFW recursion~\cite{Britto:2004ap,Britto:2005fq}.
We here address the case of general
helicity configurations using off-shell recursion.

The goal of this paper is thus to develop an off-shell recursion within the
framework of color multiplet bases, enabling efficient computation of
all-gluon amplitudes directly in an orthogonal basis.
This provides a
new route to the computation of high-multiplicity QCD matrix
elements. Unlike previous recursion schemes which target
color-ordered or dual amplitudes, the present framework is explicitly
designed to directly compute color-projected amplitudes in the
orthogonal multiplet basis.

The remainder of this paper is organized as follows. In the next
section we give an introduction to multiplet bases and color decomposition relying on 6$j$s and 3$j$s.
In section~\ref{sec:Recursion}
we introduce our recursion method and derive its complexity scaling
with gluon multiplicity. We conclude our work in
section~\ref{sec:conclusion}.

\section{Multiplet basis color reduction using Wigner $6j$ coefficients}
\label{sec:reduction}

To perform manipulations in color space, we will make extensive use
of the graphical birdtrack method; see ref.~\cite{Cvitanovic:2008zz} for a
comprehensive, general introduction to this method, and
e.g.~refs.~\cite{Keppeler:2017kwt,Peigne:2023iwm} for more dedicated
introductions to birdtracks for QCD. This method relies on
constructing diagrams with connected (and possibly disconnected)
lines, with each line representing a color representation. In this
paper color octets are denoted by curly lines; color
(anti-)triplets by plain lines with an arrow; and general or
unspecified representations by double lines with a representation
label. If both ends of a line are connected to other lines, we
implicitly sum over the representation indices of the line. For
example, the color octet indices in the loop of a "self-energy"
diagram are summed over, giving
\begin{equation}
  \label{eq:Ca}
  \raisebox{-0.5 cm}{\includegraphics[scale=0.5]{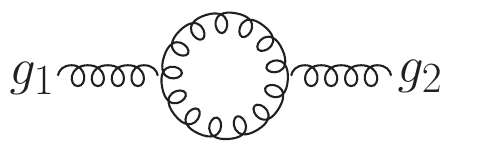}} 
  =C_{\mathrm{A}} \raisebox{-0.1 cm}{\includegraphics[scale=0.5]{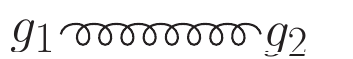}},
\end{equation}
where $C_{\mathrm{A}}$ is the Casimir operator of the adjoint
representation.  In this language, the common trace bases consist of
vectors of the form
\begin{equation}
  \label{eq:Trace basis}
  \raisebox{-0.5 cm}{\includegraphics[scale=0.5]{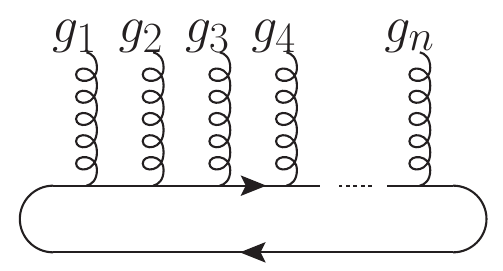}} =\tr[t^{g_1} t^{g_2} \hdots t^{g_n}]\;.
\end{equation}
The color vector space for a tree-level $n$-gluon amplitude is
spanned by all $(n-1)!$ orders in which the gluons can connect to the
triplet.  Although these spanning sets are conceptually
straightforward and enable elegant recursion
relations~\cite{Berends:1987me,Britto:2004ap,Britto:2005fq}, the
associated vectors are orthogonal only when the number of colors goes
to infinity. Since the size of the basis grows factorially with the
number of gluons, the amplitude squaring step scales with the square
of a factorial. This computational complexity motivates the
development of orthogonal multiplet bases based on representation
theory.

\subsection{Multiplet bases}
\label{sec:multiplet bases}
Multiplet bases achieve orthogonality by grouping sets of particles
into different orthogonal representations. While this is reasonably
straightforward as long as only fundamental and anti-fundamental
representations are involved, the treatment for processes involving
gluons is non-trivial and was first addressed for four
gluons~\cite{MacFarlane:1968vc,Butera:1979na}, then for
five~\cite{Sjodahl:2008fz}, and only later for a general number of
gluons~\cite{Keppeler:2012ih}. The basis vectors are of the form
\begin{equation}
  \label{eq:MSSKvector}
  \Vec^{\alpha} \equiv
  \raisebox{-0.5 \height}{\includegraphics[scale=0.4]{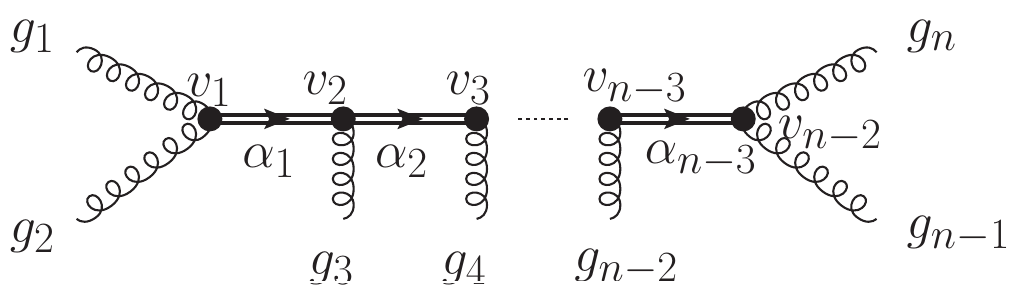}}.
\end{equation}
Contrary to trace bases, \eqref{eq:Trace basis}, the
gluon order in these multiplet basis vectors is fixed. Instead, the color space
is spanned by considering all possible representations
$\alpha_1,...,\alpha_{n-3}$ (and in general the vertices $v_1,...,v_{n-2}$)
in~\eqref{eq:MSSKvector}. Changing the gluon order would give another basis.

These bases have the advantage that they are
orthogonal. Furthermore,  an upper
bound for the number of basis vectors is given by $8^{n-3}$, since
there are $n-3$ general representations in the vector and
the number of representations in $8\otimes \alpha_i$
is limited by 8 for any representation $\alpha_i$~\cite{Keppeler:2012ih}.
This number already includes the allowed vertices, $v_1,\ldots,v_{n-2}$.
In
practice, there are significantly fewer vectors, cf.~ref.~\cite{Sjodahl:2015qoa}.
The bases can also be extended
to involve quarks~\cite{Keppeler:2012ih,Sjodahl:2018cca}, and remain
valid to any order in perturbation theory.

In general, any QCD color structure with $n$ external gluons can be
reduced into the above basis by computing the inner products of these
color structures with the basis vectors. This process is
computationally expensive when performed Feynman diagram by Feynman
diagram, which motivates the use of recursive methods. Before
introducing the off-shell recursion, we first summarize, in the next
section, the birdtrack rules that we use to manipulate the color
structures.

\subsection{Diagrammatic manipulation of color structures}
\label{sec:colour reduction}
Below we spell out the rules that we will need to manipulate the
color structures we will encounter. Examples of color reduction in
terms of $6j$s are found in refs.~\cite{Sjodahl:2015qoa,Du:2015apa}.

The first rule, the "self-energy" color relation, can be written as
\begin{equation}
  \label{eq:self-energy}
  \raisebox{-0.8 cm}{\includegraphics[scale=0.5]{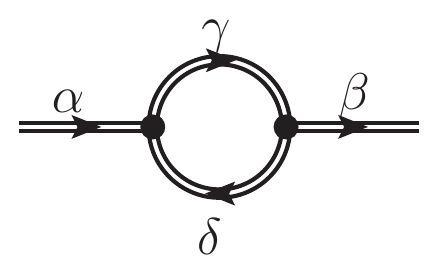}} 
  =\underbrace{\frac{\quad\includegraphics[scale=0.5]{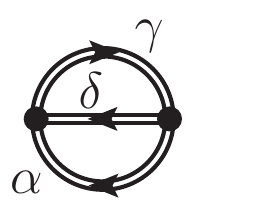}}{d_\alpha}}_{\text{real number}}
  \raisebox{-0.1 cm}{\includegraphics[scale=0.5]{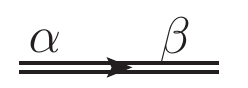}}
  \ ,
\end{equation}
for a general representation $\alpha$, which by color conservation equals $\beta$,
and for general representations $\gamma$ and $\delta$. In the numerator, the fully connected graph
is a Wigner $3j$ coefficient. The $d_{\alpha}$ in the denominator is the dimension of the representation $\alpha$.
The $3j$s can be normalized to 1 by absorbing a factor into
the vertices, but we will keep a general normalization in this work.
The Casimir operator relation in \eqref{eq:Ca} is, of course, just a special case
of \eqref{eq:self-energy}.

Triangle color loops can similarly be reduced with the
"vertex-correction relation",
\begin{equation}
  \label{eq:vertex-correction}
  \raisebox{-0.8 cm}{
    \includegraphics[scale=0.47]{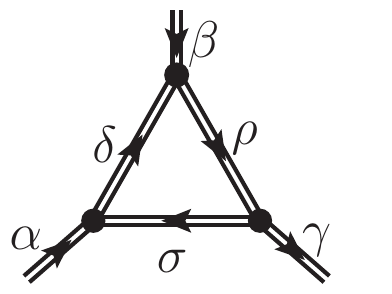}}
    \; =
     \sum_v
     \;
     \underbrace{\frac{1}{\raisebox{-0.8 cm}{
           \includegraphics[scale=0.5]{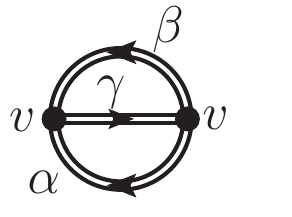}}}
     \;
     \underbrace{
     \raisebox{-0.8 cm}{\includegraphics[scale=0.47]{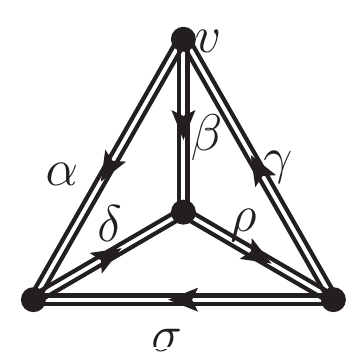}}  
     }_{\text{Wigner $6j$}}}_{\text{real number}}
     \hspace{2mm}
     \raisebox{-0.8 cm}{\includegraphics[scale=0.47]{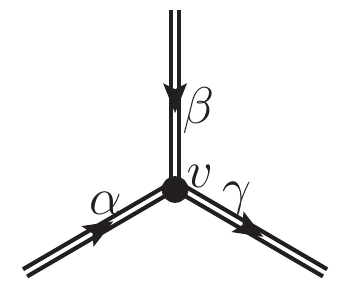}}
     \ .
\end{equation}
Here the $3j$ symbols appear in the denominators, while the numerators
contain Wigner $6j$ symbols involving six fully contracted
representations.  The sum runs over all vertices $v$ appearing in
$\alpha\otimes\beta$, resulting in a representation $\gamma$. In this
work, where at least one of the lines in any vertex is a color octet,
there are at most two possible vertices~\cite{Keppeler:2012ih}.  The
Wigner $6j$ symbols are---as any fully contracted graph---just real
numbers, which can be calculated once and for all.

Boxes, pentagons, etc.~can always be reduced to triangles using the completeness relation,
\begin{equation}
  \label{eq:completeness-relation}
   \raisebox{-0.9 cm}{\includegraphics[scale=1]{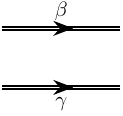}}
  \; =
  \sum_{\delta}
  \underbrace{\frac{d_{\delta}}{\includegraphics[scale=1]{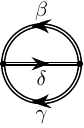}}}_\text{real number}
  \;
   \raisebox{-0.9 cm}{\includegraphics[scale=1]{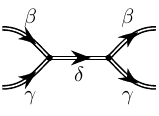}}
  \ ,
\end{equation} 
i.e., the direct product of any two representations (here $\beta$ and
$\gamma$) can be rewritten as a direct sum over representations (here
$\delta$).  By judiciously inserting completeness
relations, loops with many lines can be reduced to loops with
  fewer, which eventually can be removed by applying 
\eqref{eq:vertex-correction}.

The above rules can be combined to derive a rule for moving
a gluon that is attached to a chain of general representations, beyond an adjacent gluon, i.e.,
to "swap" the places of two gluons $g_1$ and $g_2$ in
\begin{equation}
  \label{eq:gluon move 1}
  \raisebox{-0.8 cm}{\includegraphics[scale=0.47]{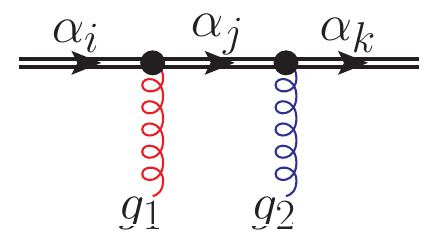}} \;.
\end{equation}
To derive this relation, we insert a completeness relation between the gluon
$g_1$
and the representation $\alpha_k$. We then use the vertex-correction relation,
\eqref{eq:vertex-correction}, to simplify the expression,
\begin{eqnarray}
  \label{eq:gluon moved}
  \raisebox{-0.8 cm}{\includegraphics[scale=0.47]{jaxodraw/GluonToMove}}
  &=&
  \sum_{\gamma} \frac{d_\gamma}{\raisebox{-0.8 cm}{\includegraphics[scale=0.47]{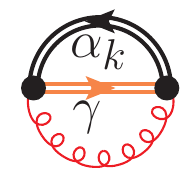}}}
  \raisebox{-1.2 cm}{\includegraphics[scale=0.47]{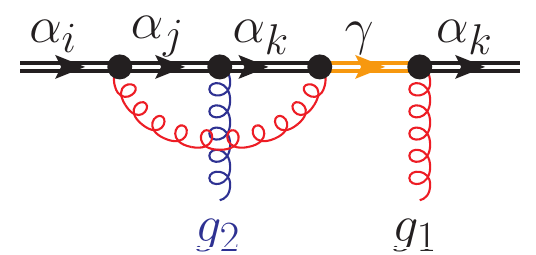}} \\
  &=&   \sum_{\gamma,\, v}
  \underbrace{\frac{d_\gamma}{\raisebox{-0.8 cm}{\includegraphics[scale=0.47]{jaxodraw/Completeness3j.pdf}}}
  \frac{\raisebox{-1.2 cm}{\includegraphics[scale=0.47]{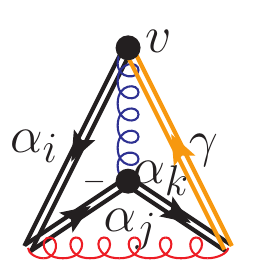}}}
       {\raisebox{-0.8 cm}{\includegraphics[scale=0.47]{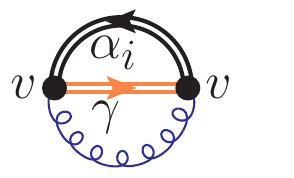}}}}_{\equiv \W_\gamma,\text{ a real number}}
  \raisebox{-1.2 cm}{\includegraphics[scale=0.47]{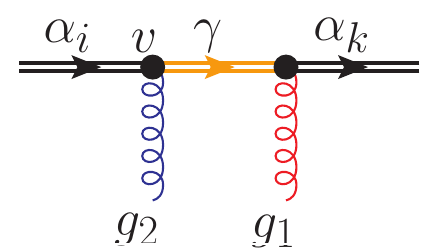}}.\nonumber
\end{eqnarray}
Thus the swap of two (neighboring) gluons introduces a
sum over representations
  $\gamma$ where each term is associated with a factor $\W_\gamma$ (in
which the other representation labels are suppressed).  A subtlety in
the above expression is the minus sign in the middle vertex of the
$6j$. This indicates that the representations entering the vertex
should be read in reverse order. Often, this is irrelevant, but
occasionally---as for the triple gluon vertex---it yields a minus
sign.

With these rules any color structure appearing in the computation of
QCD amplitudes can be projected onto the multiplet bases vectors using
$3j$ and $6j$ symbols. What remains to be addressed is the
evaluation of these symbols themselves. However,
recent work~\cite{Alcock-Zeilinger:2022hrk,Keppeler:2023msu} outlines
a novel, efficient method---since it employs fully closed or recursive
derivations without relying on the explicit calculation of Clebsch-Gordan
coefficients or symmetrizers and antisymmetrizers~\cite{Alex_2011,Dytrych:2021qwe}---to
compute all the required symbols.

\section{Off-shell recursion}
\label{sec:Recursion}

The core idea behind off-shell recursive methods is to exploit the
fact that off-shell currents can be built by summing over Feynman
diagrams where a subset of internal propagators and vertices is
shared. This significantly reduces computational redundancy compared
to evaluating all Feynman diagrams independently. This observation has
been formalized in ref.~\cite{Berends:1987me}, by introducing
off-shell currents: gauge-invariant building blocks that recursively
encode the contributions of multiple external legs attached to an
internal off-shell line. These currents are constructed by combining
lower-multiplicity currents through three- and four-gluon vertices, capturing
the full color-ordered amplitude once contracted with a final external polarization
vector.

The off-shell recursion is remarkably simple when formulated in trace
or color-ordered bases. The recursion only needs to generate planar
diagrams that respect the cyclic ordering of external legs. This
drastically reduces the number of terms generated at each recursion
step and ensures that each color-ordered amplitude corresponds
uniquely to a single color structure. As a result, the kinematic
recursion and color structures are cleanly separated, which
simplifies both the implementation and interpretation.

This simplicity does not carry over to multiplet bases. The
main complication arises because the recursion in these bases does not
preserve any fixed ordering of external gluons. While in trace-based
recursion building a current for an ordering involves combining
subcurrents that conserve the planar-ness of the contributions, for
multiplet bases there is no corresponding simple rule.
This leads to a substantial increase in the number of
required combinations. Moreover, since the
intermediate states belong to different multiplet representations,
each recursive step involves not just momentum and polarization data,
but also the evaluation of group-theoretical Wigner $3j$ and $6j$ symbols.

The development of a recursion formalism compatible with multiplet
bases thus requires a framework that can systematically handle the
non-ordered nature of the color space and the
representation-theoretic complexity introduced at each step. In the
following subsection, we present such a formalism and demonstrate how
it allows the efficient computation of off-shell currents directly in
orthogonal color bases.

Before introducing the recursion method, a subtle but important point
concerns the treatment of the four-gluon vertex. While one can include
the four-gluon Feynman rule directly into the recursion, this leads to a
combinatorial growth in the number of vertex configurations. To avoid
this, it is more efficient to express the four-gluon interaction in terms
of effective combinations of three-particle vertices by introducing an
auxiliary non-propagating anti-symmetric tensor-like
particle\footnote{Equivalently, one could upgrade the gluon propagator
to a tensor particle, where the diagonal contributions form the usual
gluon, and the off-diagonal terms the non-propagating anti-symmetric
tensor-like particle.}~\cite{Duhr:2006iq}. This "vertex-splitting"
approach not only reduces the scaling complexity, it also simplifies
our description of the recursion in multiplet bases
as we can effectively ignore the four-gluon interaction in our
discussion.

\subsection{Recursion in multiplet bases}
The basic objects in our recursive method to compute an $n$-gluon
amplitude are $k$-gluon currents, with $k<n$. By construction, we
define an $(n-1)$-gluon current $\cur_{(n-1)}^{\mu}$ (by momentum
conservation this is an on-shell current), such that when contracted
with the polarization vector for gluon $a_n$, it yields the
coefficient $\bvec$ multiplying one of the basis vectors of the
$n$-gluon multiplet basis,
\begin{multline}
  \bvec(a_1,\ldots,a_n; \alpha_1,\ldots,\alpha_{n-3}; v_1,\ldots,v_{n-2})=\\
  \cur^{\mu}_{(n-1)}(a_1,\ldots,a_{n-1}; \alpha_1,\ldots,\alpha_{n-3}; v_1,\ldots,v_{n-2})\epsilon_{\mu}(a_n).
\end{multline}
Here and in the following, with some abuse of notation, we use $a_i$
to identify one of the gluons through its momentum and helicity. That is,
$\epsilon_{\mu}(a_n)\equiv \epsilon_{\mu}(p_{a_n},\lambda_{a_n})$ is the
polarization vector for the gluon with momentum $p_{a_n}$ and helicity
$\lambda_{a_n}$.
In general, any $k$-gluon current depends on the $k$ external gluons,
 denoted by $\{a_i\}$, the $k-2$
color representation labels $\{\alpha_i\}$, and the $k-1$ vertices
$\{v_i\}$. The simplest current---a $1$-gluon
current---is just the polarization vector of the corresponding gluon,
\begin{equation}
\cur^{\mu}_{(1)}(a_1)=\epsilon^{\mu}(a_1).
\end{equation}

\begin{figure}[t]
  \begin{center}
  \includegraphics[scale=0.45]{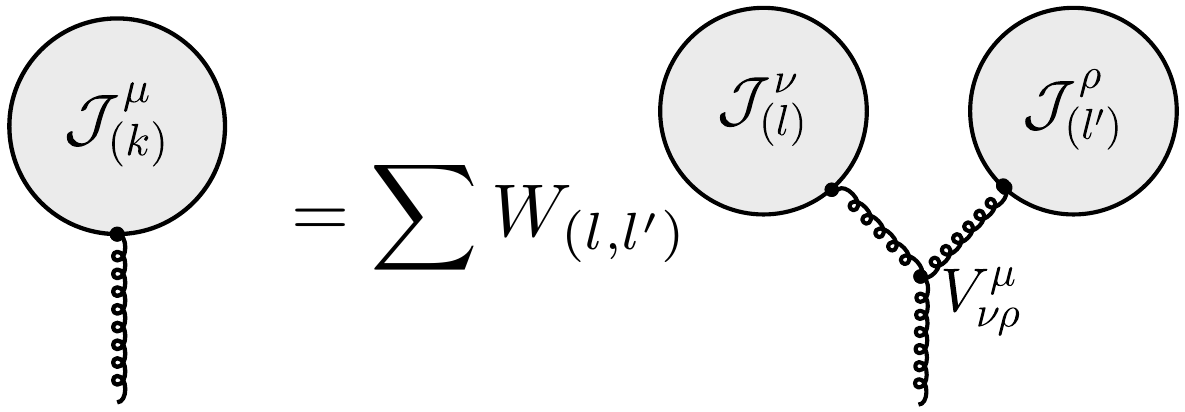}
  \end{center}
  \caption{Graphical representation of the recursion formula~\eqref{recursion}.}
  \label{fig:recursion}
\end{figure}

Any current with $1<k<n$ can be computed recursively from lower-point currents, see \figref{fig:recursion},
\begin{align}
  \cur^{\mu}_{(k)}(\{a,\alpha\})&=
  \sum \,W_{(l,l^{\prime})}(\{a,\alpha\};\{b,\beta\},\{b^{\prime},\beta^{\prime}\})\,\cur^{\nu}_{(l)}(\{b,\beta\})\,\cur^{\rho}_{(l^{\prime})}(\{b^{\prime},\beta^{\prime}\})\,V^{\mu}_{\nu\rho}(p_{\{b\}},p_{\{b^{\prime}\}})\nonumber\\
  \label{recursion}
  &\equiv\sum \,W_{(l,l^{\prime})}(\{a,\alpha\};\{b,\beta\},\{b^{\prime},\beta^{\prime}\})\,\mathcal{K}^{\mu}_{(l,l^\prime)}(\{b,\beta\},\{b^{\prime},\beta^{\prime}\}),
\end{align}
where we have introduced the short-hand notation
\begin{equation}
  \label{eq:new current}
  \cur^{\mu}_{(k)}(\{a,\alpha\})\equiv
\cur^{\mu}_{(k)}(a_1,\ldots,a_{k}; \alpha_1,\ldots,\alpha_{k-2};
v_1,\ldots,v_{k-1}) 
\end{equation}
with the dependence on the vertex types $\{v_i\}$ understood,
and, in the second line introduced the kinematics object
$\mathcal{K}^{\mu}_{(l,l^\prime)} \equiv
\cur^{\nu}_{(l)}\cur^{\rho}_{(l^{\prime})}V^{\mu}_{\nu\rho}$.  In this
equation $V^{\mu}_{\nu\rho}(p_{\{b\}},p_{\{b^{\prime}\}})$ encodes the
color-stripped three-gluon Feynman rule and, for $k<n-1$, includes
the new gluon propagator. It depends only on the total momenta of
the gluons entering the subcurrents $\cur^{\nu}_{(l)}$ and
$\cur^{\rho}_{(l^\prime)}$, respectively\footnote{See the end of the
previous section for a discussion of the four-gluon interaction and why
it can be ignored here.}. The color coefficients
$W_{(l,l^{\prime})}(\{a,\alpha\};\{b,\beta\},\{b^{\prime},\beta^{\prime}\})$
are constructed from the color structures of $\{b,\beta\}$ and
$\{b^{\prime},\beta^{\prime}\}$, coupling in a triple-gluon vertex to
the new off-shell gluon by projecting it onto the color structure
of $\{a,\alpha\}$. 

The sum in \eqref{recursion} runs over all non-empty partitions
of $a_1, \ldots, a_k$ into two indistinguishable, unordered subsets,
hence $l+l^{\prime}=k$, as well as over all possible representations
and vertex types allowed in $\cur^{\nu}_{(l)}$ and
$\cur^{\rho}_{(l^{\prime})}$ that, when combined, give a contribution
to the current $\cur^{\mu}_{(k)}(\{a,\alpha\})$. Note that we require
only two \emph{unordered} subsets, since each basis vector (and thus
also each current) uses a fixed, canonical ordering of the gluons,
$a_1<a_2<\ldots<a_k$, and similarly for the $b_i$'s and
$b^{\prime}_i$'s.

To estimate the computational complexity of evaluating a color-summed
$n$-gluon matrix element, observe that at most
\begin{equation}
  \label{numberofcurrents}
x_k=\binom{n-1}{k}8^{k-2}
\end{equation}
different $k$-currents are needed in the computation of the
color-summed squared amplitude. Here the binomial factor selects the
$k$ gluons out of the possible $n-1$, and the factor\footnote{Strictly
speaking, this factor should be replaced by $\max(8^{k-2},1)$ to take
into account that for $k=1$ there are exactly $n-1$ different
currents.} $8^{k-2}$ accounts for the number of allowed
representations and associated vertex types, see section~\ref{sec:multiplet bases}. From the
recursive structure in~\eqref{recursion}, the number of terms in the
sum for each $k$-current is limited to
\begin{equation}
  \label{naive_number_of_coefficients_per_current}
  m_k=\frac{1}{2}\sum_{i=1}^{k-1}\binom{k}{i}8^{i-2}\, 8^{k-i-2}=\frac{1}{2}8^{k-4}\sum_{i=1}^{k-1}\binom{k}{i},
\end{equation}
where the factor $\frac{1}{2}\sum_{i=1}^{k-1}\binom{k}{i}$ denotes the
number of ways $k$ gluons can be split into two indistinguishable,
non-empty sets, and $8^{i-2}$ and $8^{k-i-2}$ denote the (overestimates
for the) number of color representations in those two sets.  Hence, in
total the number of operations needed to compute all basis vectors is
bounded by
\begin{equation}
  \label{naivescaling}
\sum_{k=1}^{n-1}x_k\,m_k=\mathcal{O}(129^n).
\end{equation}
Finally, since the multiplet basis is orthogonal, constructing the
full color-summed $n$-gluon squared amplitude from the basis vectors
requires only a negligible additional cost of $\mathcal{O}(8^{n-3})$
operations. Therefore, using the multiplet bases, the computation of
$n$-gluon color-summed matrix elements can be performed with
algorithms of at most \emph{exponential} complexity.

Although the scaling found in~\eqref{naivescaling} is only
exponential, the base of the exponential is so large that for any
reasonable number of gluons the number of operations required exceeds
the $\mathcal{O}(((n-1)!)^2)$ operations needed to compute
color-summed squared amplitudes using the trace
basis. However, our derivation of~\eqref{naivescaling} is rather
naive. In particular, it assumes that all $W_{(l,l^{\prime})}$ on the
right-hand side of~\eqref{recursion} are independent and their values
must be computed independently for every term in the sum. In practice
this is not the case. By carefully rewriting the sum, one can maximize
the interdepence among the $W_{(l,l^{\prime})}$ to reduce the overall
computational complexity.  Moreover, since all possible currents must be
computed, and many of the intermediate results contribute
to more than one current, substantial savings can be achieved through
caching and reuse of partial computations across different
currents. We will describe this optimization procedure in
the next section. The proposed algorithm reduces the computational
complexity by almost a factor $8^{n}$, leading to the improved
scaling
\begin{equation}
  \label{improvement}
  \mathcal{O}(129^n) \quad \to \quad \mathcal{O}(17^n).
\end{equation}
Consequently, for even moderately large values of $n$
the multiplet basis has already the potential to become more efficient
than, e.g., the trace basis for computing color-summed matrix
elements.

\subsection{The color coefficients and the optimized algorithm}
The color coefficients
$W_{(l,l^{\prime})}(\{a,\alpha\};\{b,\beta\},\{b^{\prime},\beta^{\prime}\})$
can be obtained by calculating the scalar product between the basis
vector corresponding to the structure $\{a,\alpha\}$ and the vector
obtained by combining $\{b,\beta\}$ and
$\{b^{\prime},\beta^{\prime}\}$ through a three-gluon vertex. To
achieve an \textit{efficient} decomposition, however, we will
introduce an algorithm to determine $W_{(l,l^{\prime})}$ in
terms of the Wigner $6j$ and $3j$ symbols.

We start from the color configuration corresponding
to
$\mathcal{K}^{\mu}_{(l,l^\prime)}(\{b,\beta\},\{b^{\prime},\beta^{\prime}\})$. In
birdtrack notation, and omitting the vertex labels, this is given by
\begin{equation}
  \label{one_chain}
  \raisebox{-0.5 \height}{\includegraphics[scale=0.45]{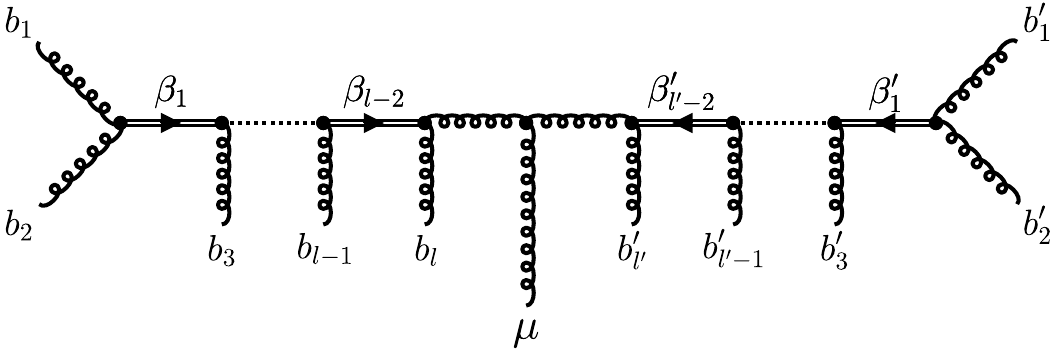}},
\end{equation}
with the off-shell gluon labeled by $\mu$. (We have ignored potential minus
signs coming from the fact that the representations entering the
vertices on the right of the gluon $\mu$ should be read in reverse
order).

We have to find the projection of this configuration onto the color
basis vector relevant to $\cur^{\mu}_{(k)}(\{a,\alpha\})$. This can be
obtained by manipulating the diagram
in~\eqref{one_chain}, until all the gluons are in canonical order by
repeatedly applying the swapping rule, \eqref{eq:gluon moved}. A single swap of two
adjacent gluons (highlighted in red), results in a sum over all
possible intermediate color representations $\gamma_l$,
\begin{equation}
  \label{one_swap}
  \textrm{\eqrefeq{one_chain}} = \sum_{\gamma_l}\W_{\gamma_l}\raisebox{-0.60 \height}{\includegraphics[scale=0.45]{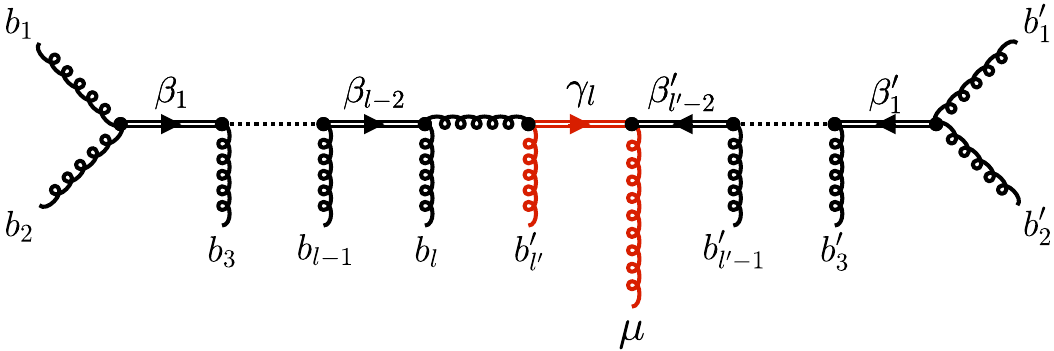}},
\end{equation}
where $\W_{\gamma_l}$ is given by \eqref{eq:gluon
  moved}, and depends on the Wigner $3j$ and $6j$ symbols.

After two swaps we get
\begin{equation}
  \label{two_swaps}
  \textrm{\eqrefeq{one_chain}} = \sum_{\gamma_l,\gamma_{l-1}}\W_{\gamma_l}\W_{\gamma_{l-1}}\raisebox{-0.60 \height}{\includegraphics[scale=0.45]{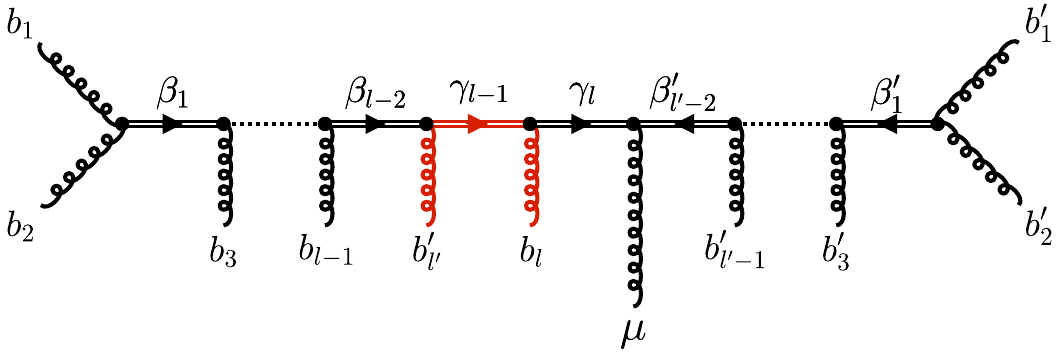}},
\end{equation}
and after three swaps
\begin{equation}
  \label{three_swaps}
  \textrm{\eqrefeq{one_chain}} = \sum_{\gamma_l, \gamma_{l-1}, \gamma_{l-2}}\W_{\gamma_l}\W_{\gamma_{l-1}}\W_{\gamma_{l-2}}\raisebox{-0.60 \height}{\includegraphics[scale=0.45]{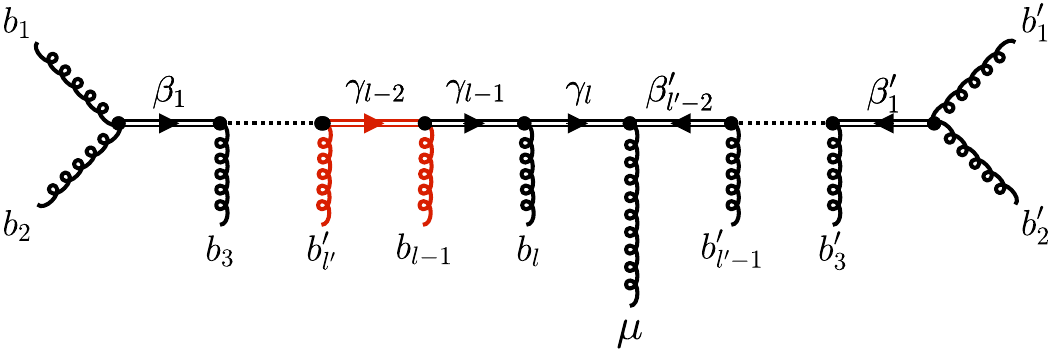}},
\end{equation}
etc. Thus, an increasing number of sums over intermediate
representations is introduced, together with the corresponding $\W_i$
factors. When all the swaps are completed, that is, \emph{all} the gluons
are in the canonical order, $a_1<a_2<\ldots<a_k$ and with the
off-shell gluon $\mu$ in the rightmost position,
\begin{equation}
  \label{all_swaps}
  \textrm{\eqrefeq{one_chain}} = \sum_{\{\gamma\}\ldots\{\alpha^{\prime}\}}\W_{\{\gamma\}}\ldots \W_{\{\alpha^{\prime}\}}\raisebox{-0.5 \height}{\includegraphics[scale=0.25]{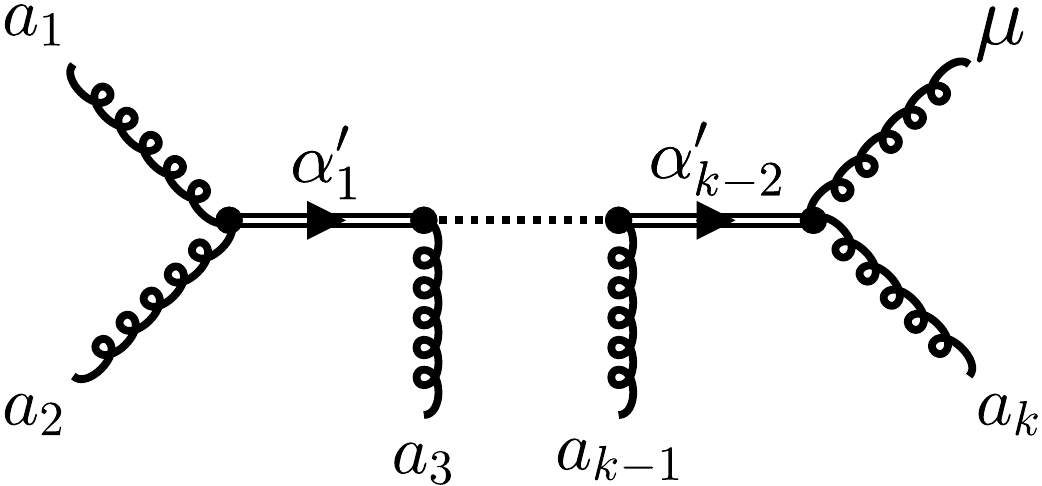}},
\end{equation}
the color coefficients are given by
\begin{equation}
  \label{color-coeff}
  W_{(l,l^{\prime})}(\{a,\alpha\};\{b,\beta\},\{b^{\prime},\beta^{\prime}\})=
  \sum_{\{\gamma\}\ldots\{\alpha^{\prime}\}}
  \W_{\{\gamma\}}\ldots \W_{\{\alpha^{\prime}\}}\,
  \delta_{\alpha^{\prime}_1\alpha_1}\ldots\delta_{\alpha^{\prime}_{k-2}\alpha_{k-2}},
\end{equation}
with $\delta_{ij}$ the usual Kronecker deltas.

It is clear from this algorithm that the color coefficients,
\eqref{color-coeff}, are not independent across all possible
$\{a,\alpha\}$, $\{b,\beta\}$, $\{b^{\prime},\beta^{\prime}\}$, and
partitions of $k$ into $l$ and $l^{\prime}$. There are essentially two main ways one can
encounter identical (intermediate) birdtrack configurations. The first
is that after a swap a representation gets replaced by a sum over
representations, and, as such, all possible values of the original
representation can be treated together. For example, after the swap of
the gluons labeled $b'_{l'}$ and $b_{l-1}$ in \eqref{three_swaps}, the
representation $\beta_{l-2}$ is irrelevant, showing up just in the
coefficient $\W_{\gamma_{l-2}}$. Thus, at this stage contributions
starting with different $\beta_{l-2}$ can be combined. The second is
that, once a gluon has reached its final position---suppose, e.g.,
that $b'_{l'}$ in \eqref{three_swaps} is in its correct
location---that configuration is (in the case both $\gamma_l$ and
$\beta^{\prime}_{l^{\prime}-2}$ are color octets) identical to another
possible starting configuration, cf. \eqref{one_chain}, for a
different partition of the $k$ gluons into two sets, i.e., the one
where $b'_{l'}$ is already part of the set $\{b\}$.  Hence, different starting color
representations $\{b,\beta\}$ and $\{b^{\prime},\beta^{\prime}\}$, as
well as different partitions of the $k$ gluons into two sets with $l$
and $l^{\prime}$ gluons, can lead to the same intermediate
birdtrack diagram.
This means that, at each such stage in the
procedure, the corresponding kinematic factors
$\mathcal{K}^{\mu}_{(l,l^\prime)}$---multiplied by the
already-included product of factors $\W_i$---for any
configurations that lead to the same intermediate state can be summed
together.  The remaining $\W_i$ factors introduced after that point
only need to multiply the already-summed combination of kinematic
factors.

Furthermore, because in the final step we need to introduce Kronecker
deltas, see~\eqref{color-coeff}, our algorithm effectively obtains all
possible representations for $\alpha_1,\ldots,\alpha_{k-2}$ (and vertex
types) in one go.

The above considerations result in a greatly reduced number of
operations needed to obtain all possible $k$-gluon currents. Instead
of $x_k\,m_k$, see eqs.~(\ref{numberofcurrents}) and
(\ref{naive_number_of_coefficients_per_current}), we only need
\begin{equation}
  8^{k-2}u_k
\end{equation}
operations, where the factor $8^{k-2}$ counts the number of possible
representations for a given order of the gluons and $u_k$ is the
number of unique orderings of the gluons one encounters within the
swapping procedure. The latter scales like $2^k$ for large $k$. One
way to see this is to notice that the number of partitions of $k$
objects into two unsorted sets scales like $\mathcal{O}(2^{k})$. For
each of those pairs of sets, a bubble sort can order the elements in
$\mathcal{O}(k^2)$ operations\footnote{In practical implementations,
when dealing with moderately large values of $k$, this is one place
where one can easily find optimizations. For example, by considering
currents with both canonical and reverse ordering (at a price of
needing to compute twice as many currents) and performing a reverse
bubble sort to order the gluons in the birdtrack diagrams, the value
for $u_k$ would be equal to $k\,2^{k-1}-1$.}, which is negligible
compared to $2^k$ for large $k$. Therefore, in total, we find that the
computational complexity to compute $n$-gluon color-summed squared
matrix elements scales as
\begin{equation}
  \label{optimised_scaling}
\sum_{k=1}^{n-1}\binom{n-1}{k}8^{k-2}2^k=\mathcal{O}(17^n).
\end{equation}

\section{Conclusion and outlook}
\label{sec:conclusion}
We have developed an off-shell recursive formalism for computing
tree-level QCD gluon amplitudes directly in color multiplet
bases. Unlike conventional recursion relations, which target dual or
color-ordered amplitudes, used in non-orthogonal bases, our method
constructs the full amplitude projected onto an orthogonal basis of
color representations.

We have shown that using Wigner $6j$
coefficients for color space, it is possible to perform such
recursions entirely within multiplet bases. In these bases, the
color-summed squaring step becomes trivial due to orthogonality, and
the challenge shifts instead to the efficient construction of
off-shell currents. By developing a recursive algorithm for multiplet
currents, and carefully analyzing its scaling, we demonstrated that
a naive exponential complexity $\mathcal{O}(129^n)$ can be significantly
reduced to $\mathcal{O}(17^n)$ through caching and partial summation
of intermediate subcurrents. This improved scaling indicates that for
computing color-summed squared amplitudes, the multiplet basis can
outperform all commonly used bases, such as trace, color-flow and
adjoint---which are all of a squared factorial complexity---already
for a moderate number of gluons, making it a
competitive and viable framework for high-multiplicity processes in
QCD.

A natural next
step would be the implementation of this algorithm in a Monte Carlo
event generator framework to study the exact complexity of the
algorithm for moderately large values of $n$.
The extension to tree-level processes including
quarks~\cite{Sjodahl:2018cca} exactly parallels the all-gluon case
presented here. The main difference is a smaller sum over intermediate
representations when a gluon is exchanged for a quark in the swapping
rule, and therefore a better scaling with $n$. 

For phenomenology, it will be important to address the assignment of
leading-color flows to generated events for correct post-processing
such as parton showering. Recent work on two-step event
generation~\cite{Frederix:2024uvy} provides a natural solution to this
potential problem with using multiplet bases.

\section*{Acknowledgments}
R.F.~is supported by the Swedish Research Council under contract
number 202004423.

\bibliographystyle{JHEP}  
\bibliography{refs} 

\end{document}